\documentclass[preprint,12pt]{elsarticle}
\usepackage{amssymb}
\usepackage{url}
\makeatletter
\def\ps@pprintTitle{%
	\let\@oddhead\@empty
	\let\@evenhead\@empty
	\let\@oddfoot\@empty
	\let\@evenfoot\@oddfoot
}
\makeatother
\begin{document}

\begin{frontmatter}

\title{Small sample measurements at the low energy muon facility of Paul Scherrer Institute}

\author[inst1]{Xiaojie Ni\corref{cor1}}
\cortext[cor1]{Corresponding author}
\ead{xiaojie.ni@psi.ch}
\affiliation[inst1]{organization={Laboratory for Muon Spin Spectroscopy, Paul Scherrer Institute},
            city={Villigen PSI},
            postcode={5232}, 
            country={Switzerland}}

\author[inst1,inst2,inst3]{Luping Zhou}

\affiliation[inst2]{organization={Institute of High Energy Physics, CAS},
            city={Beijing},
            postcode={100049}, 
            country={China}}
            
\affiliation[inst3]{organization={University of Chinese Academy of Sciences, CAS},
            city={Beijing},
            postcode={100049}, 
            country={China}}

\author[inst1,inst4]{Maria Mendes Martins}

\affiliation[inst4]{organization={Advanced Power Semiconductor Laboratory, ETH Zurich},
            city={Zurich},
            postcode={8092}, 
            country={Switzerland}}  
            
\author[inst1]{Zaher Salman}
\author[inst1]{Andreas Suter}
\author[inst1]{Thomas Prokscha\corref{cor2}}
\cortext[cor2]{Corresponding author}
\ead{thomas.prokscha@psi.ch}

\begin{abstract}
The low energy muon spin rotation  spectroscopy (LE-$\mu$SR) is primarily used to investigate thin films, surfaces, and interfaces of materials, which has matured at the Paul Scherrer Institute (PSI) and is routinely employed by users of the low energy muon (LEM) facility. However, because of the large beam spot and low implanted muons rate, LE-$\mu$SR measurements on small samples are difficult, requiring an optimal sample size of $25\times25$ mm$^2$. Recently, we have boosted our ability to measure small samples, down to $5\times5$ mm$^2$ area, by beam collimation and tuning. This achievement is crucial for the measurements of many magnetic and superconducting materials. Furthermore, we have devised a method that allows us to measure five small area samples mounted together on the same sample plate. We expect this method to further improve the efficient use of beam time at LEM.
\end{abstract}




\end{frontmatter}


\section{Introduction}


Muon Spin Rotation, Relaxation, and Resonance ($\mu$SR) is a powerful technique for studying the local magnetic and electrical properties of materials due to its inherent sensitivity to the atomic environment surrounding the muon stopping site \citep{hillier2022}. In general, the $\mu$SR technique primarily utilizes positively charged muons ($\mu$$^{+}$) produced at the surface of a muon production target to measure bulk materials. Since these so-called surface $\mu$$^{+}$s originate from the pions' decay at rest, they are almost fully polarized and have high kinetic energy, around 4 MeV. 
The LEM facility at PSI can slow down a small fraction of surface $\mu$$^{+}$s and extend the $\mu$SR technique to the investigation of thin films, heterostructures, and surface regions with tunable $\mu$$^{+}$s' energies from 1 to 30 keV \citep{morenzoni_generation_1994,morenzoni_low-energy_2000,prokscha_moderator_2001,morenzoni_nano-scale_2004,prokscha2008}. 

\begin{figure}[!htb]
	    \centering
	    \includegraphics[width=0.95\textwidth]{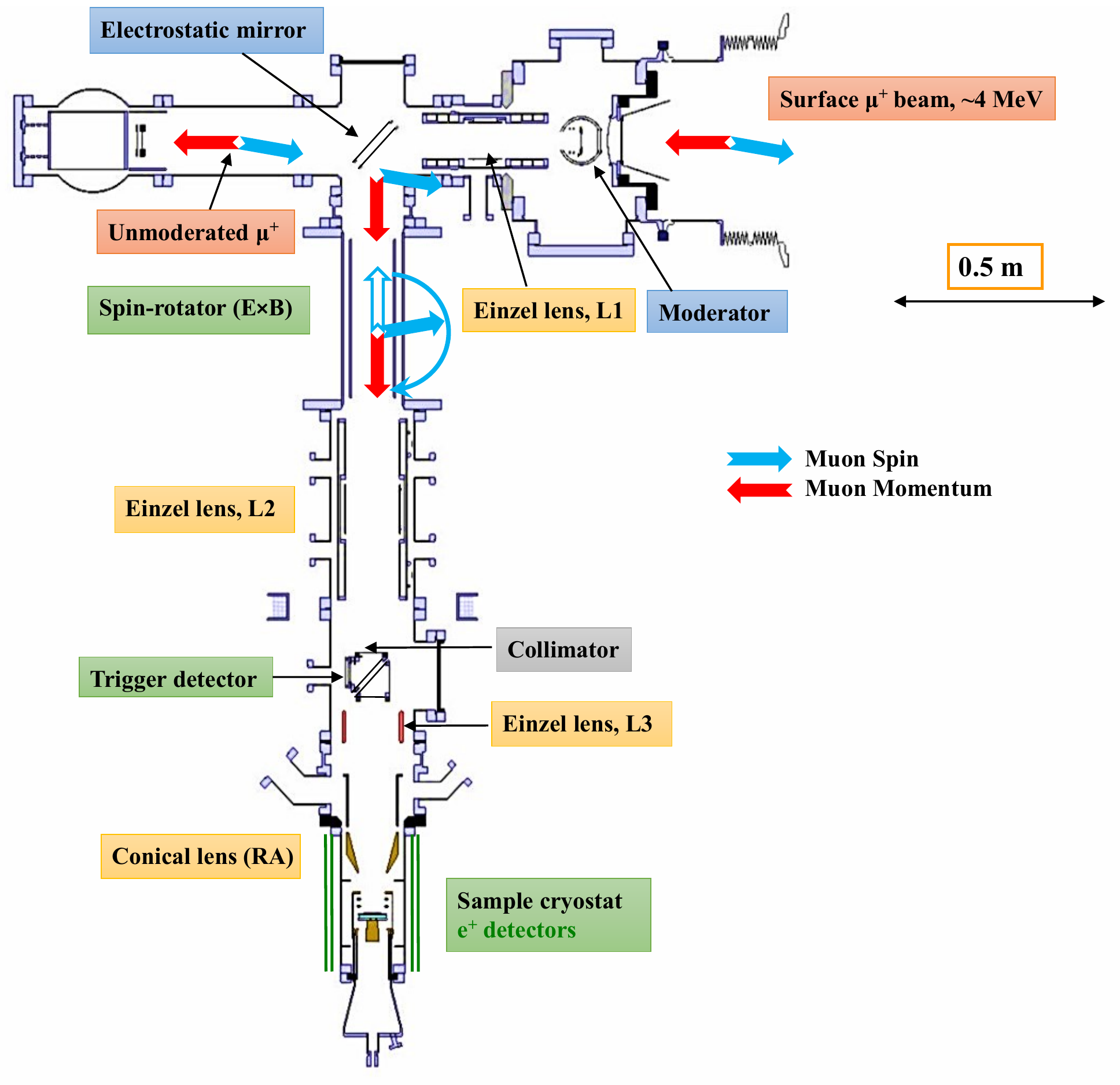}
	    \caption{Schematic of the low energy muon facility. The muon spin direction (blue arrow) at L1 is denoted
	    as +10$^{\circ}$ with respect to the momentum (red arrow) of the incoming surface muon beam. In the
	    spin-rotator, it is rotated to -10$^{\circ}$, using the spin-rotator also as a velocity filter
	    to reduce background due to protons and ions emitted at the moderator.}
	    \label{fig:LEM}
\end{figure}

As shown in Fig. \ref{fig:LEM}, the surface $\mu$$^{+}$s from the high intensity $\mu$E4 beamline impinge on a moderator to generate epithermal $\mu$$^{+}$s of approximately 15~eV. The moderator consists of a 125 $\mu$m patterned silver foil covered with around 300 nm solid Ar-N$_2$ layer whose moderation efficiency is 10$^{-5}$ -- 10$^{-4}$ \citep{prokscha_moderator_2001,morenzoni_nano-scale_2004}. After being emitted from the moderator surface, epithermal $\mu$$^{+}$s can be accelerated to different kinetic energies (up to 20 keV) by high-voltage grids near the moderator. The potential at the electrostatic mirror is the same as that at the moderator in order to bend the slow $\mu$$^{+}$ beam by 90 degrees towards the spin rotator and allow the unmoderated $\mu$$^{+}$s to pass through. The spin rotator (SR) can be set to change the $\mu$$^{+}$s' spin angle from -90 to +90 degrees with respect to the surface muon momentum direction and to reduce the background of protons and other charged particles -- originating from the moderator with different velocity -- by the $E\times B$ field or SR \citep{salman2012}. The trigger detector (TD) provides the start signal by detecting secondary electrons, which are generated from $\mu$$^{+}$s passing through a ten-nanometer-thin carbon foil \citep{morenzoni_generation_1997, khaw2015geant4}. Three Einzel lenses (L1, L2, L3) and one conical lens (RA, ring anode \cite{xiao_segmented_2017}) are used to optimally transport and focus the slow $\mu$$^{+}$ beam onto the sample plate. The potential at the sample plate can be varied between -10 and +10~kV to adjust the implantation energy of slow $\mu$$^{+}$ in the range from 1 to ~30 keV, so that they can be implanted into the sample at a mean depth range from a few nanometers to $\sim$200 nanometers.

Due to the large phase space of the $\mu$$^{+}$ beam, about 80\% of the slow $\mu$$^{+}$s stop in an area of $20\times20$ mm$^2$, and only 42\% in $10\times10$ mm$^2$. 
Measurements on $5\times5$ mm$^2$ sized samples are unfeasible because of the low fraction of the beam (16\%) stopping in this small area, making the signal-to-background ratio unacceptably low. Therefore, an optimal sample size of $25\times25$ mm$^2$ is required at the LEM facility, which restricts the measurements of some novel magnetic and superconductor materials, since they usually can only be manufactured in small sizes. In this study, we designed two collimators with diameters of 10 and 15 mm, respectively, which can be inserted in front of the carbon foil of TD by an UHV-compatible magnetic transporter. Consequently, it is possible to measure small samples with a size of $10\times10$ mm$^2$ or $5\times5$ mm$^2$ by beam collimation and tuning the beam transport with a significantly increased signal-to-background ratio and figure-of-merit. In addition, we can also measure four $10\times10$ mm$^2$ or five $5\times5$ mm$^2$ small samples mounted on the same sample plate by adjusting the high voltages of the four segments of RA (RAL, RAR, RAT, RAB) to direct the $\mu$$^{+}$ beam to these samples' positions. 

The paper is organized as follows: in Section \ref{sec:simulations}, we determine the optimum diameters of collimator by Geant4 Monte Carlo simulations \citep{sedlak2012musrsim}. Then, as shown in Section \ref{sec:Beamspot}, microchannel plate (MCP) tests were carried out to optimize the beam spots, not only at the center but also at the four corners. In Section \ref{sec:muSR}, we introduce the $\mu$SR measurements of small samples with the optimal transport settings found by the MCP, including one $10\times10$ mm$^2$ SrTiO$_3$/Nd$_{0.8}$Sr$_{0.2}$NiO$_2$/SrTiO$_3$ sample, four $10\times10$ mm$^2$ and five $5\times5$~mm$^2$ SrTiO$_3$ samples mounted on one sample plate. Finally, the summary is presented in Section \ref{sec:Summary}.

\section{Simulations}
\label{sec:simulations}
\begin{figure}[!ht]
	    \centering
	    \includegraphics[width=1.0\textwidth]{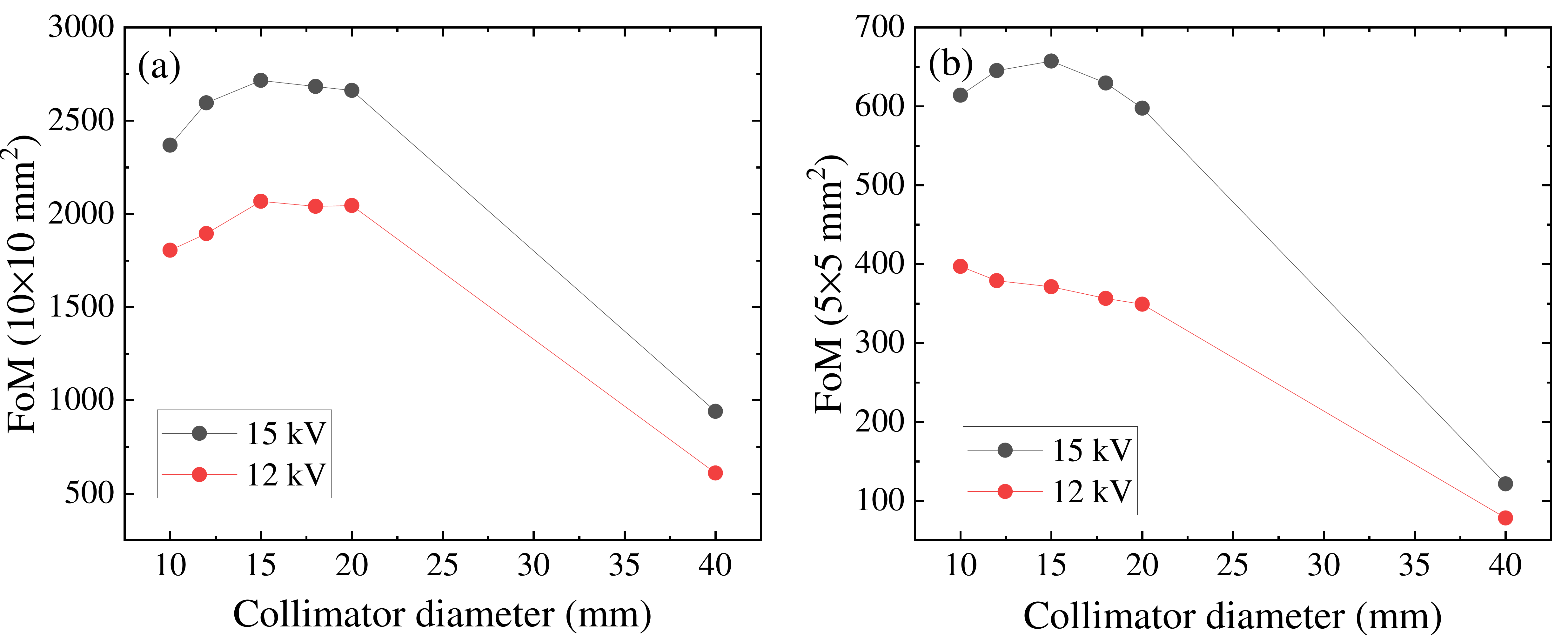}
	    \caption{Simulated figure of merit FoM as a function of collimator diameter for sample sizes of (a) $10\times 10$~mm$^2$
	    and (b) $5\times 5$~mm$^2$, respectively, at 12 and 15 kV transport settings.
	    12 and 15 kV transport settings refer to the potentials at the moderator (V$_{mod}$).
	    Based on V$_{mod}$, the high voltages of L1, L2, L3, and RA are optimized to transport the muon beam and maximize the number of muons on the sample plate.}
	    \label{fig:DifferentDiameter}
\end{figure}

To select the optimum collimator size, we performed beam transport simulations using the Geant4-based program musrSim \citep{sedlak2012musrsim}. Following the parameters of LEM components, the simulation geometry model was constructed, and all electro-magnetic field maps employed were calculated using Comsol Multiphysics\textsuperscript{\textregistered} and Opera \citep{COMSOL_web,OPERA-3D_web}. 
The muon beam with 10$^4$ initial $\mu$$^{+}$s was placed at the moderator with an initial kinetic energy of 15 eV and an FWHM of 20 eV \citep{morenzoni_low_1997}. We optimized the beam transport for different collimator sizes for 12 and 15 kV extraction voltages at the moderator, i.e. muon beam transport energies of 12 and 15 keV, respectively. These are the most frequently used transport energies in LEM experiments.
The corresponding high voltages of L1, L2, L3, and RA were adjusted in the simulation to
optimize the figure of merit (FoM). In a $\mu$SR experiment, FoM is given by the product 
$A^2\times N$, where $A$ is the measured decay asymmetry of the muons -- proportional to the polarization of the $\mu^+$ ensemble -- and $N$ is the total number of recorded muon decays.
In the LEM instrument, the samples are usually glued onto Ni-coated sample plates, where
muons missing the sample quickly depolarize in the ferromagnetic Ni layers, causing a reduction of $A$ \cite{saadaoui_zero-field_2012}. Therefore, $A$ is 
proportional to the fraction $F_{samp}$ of muons stopping in the area of the sample, and $N$ is 
proportional to the transmission $Tr$ of the muon beam up to the sample position (fraction
of muons arriving at the sample plate), which is proportional to the LEM rate, i.e. the
number of detected muon events. In the simulation, the figure of merit can then be expressed by $FoM = F_{samp}^2\times Tr$. Here, we investigate the fraction of muons stopping in areas of
$10\times10$ mm$^2$ or $5\times5$~mm$^2$ and the transmission of muons to the sample region
as a function of collimator size.
%
%
As shown in Fig. \ref{fig:DifferentDiameter}, the FoMs of $10\times10$ mm$^2$ and $5\times5$ mm$^2$ areas with collimators are significantly higher than those without collimator. The collimator with a diameter of 40 mm represents reference values without a collimator since the diameter of the carbon foil frame of TD is 40 mm. At V$_{mod}$~=~15 and 12 kV, the 15-mm collimator performs best for $10\times10$ mm$^2$ area sample measurements. A 15-mm collimator is best for measuring $5\times5$ mm$^2$  area samples at V$_{mod}$ = 15 kV, while a 10-mm collimator is best at V$_{mod}$ = 12 kV.

\section{Beam spot measurements}
\label{sec:Beamspot}
\begin{figure}[!ht]
	    \centering
	    \includegraphics[width=0.9\textwidth]{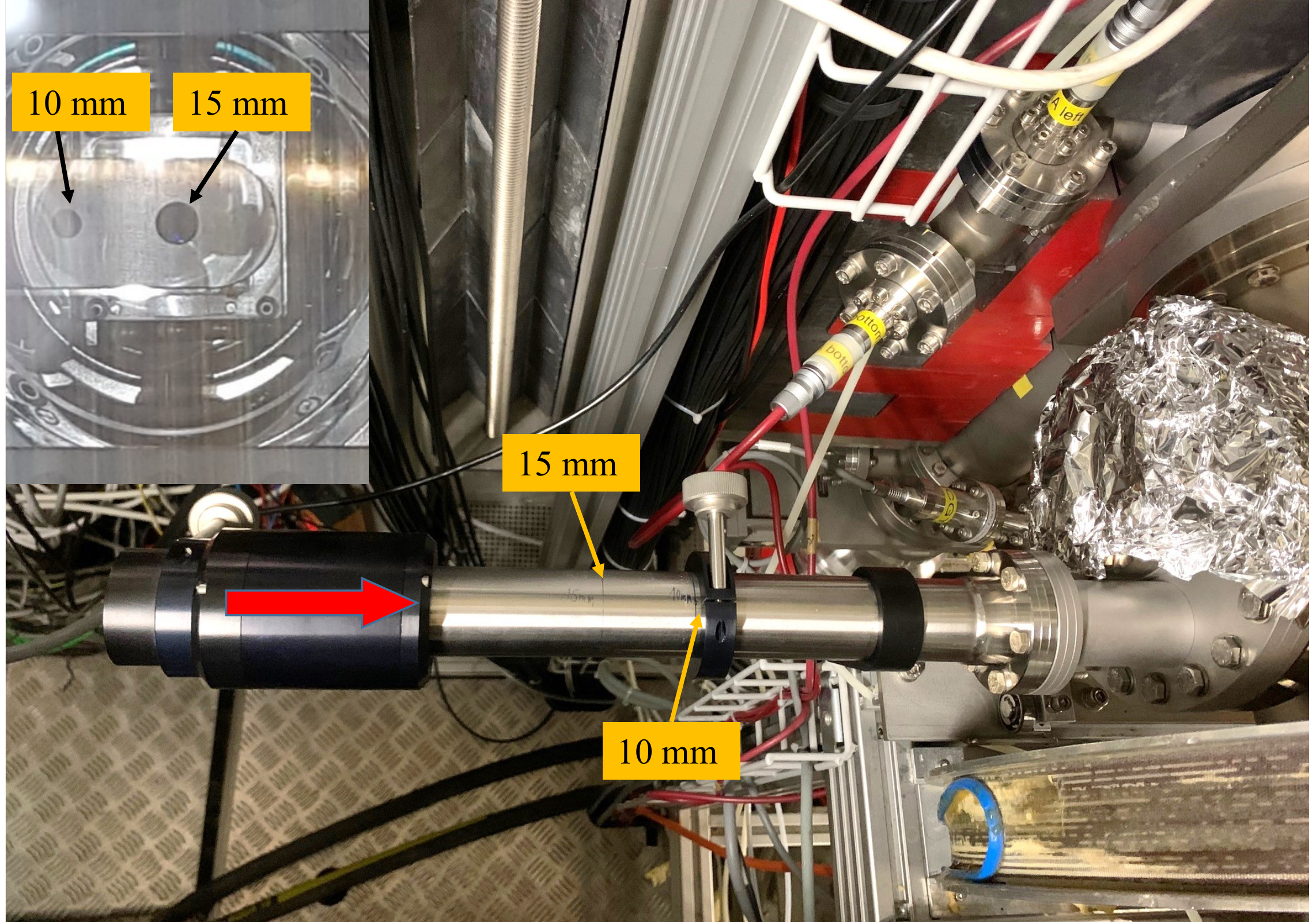}
	    \caption{The magnetic transporter with 10- and 15-mm-diameter collimators attached to its insert on its vacuum side. Marking the collimator size on the transporter enables accurate positions. The inset picture was taken from upstream of TD at a viewport mounted at the vacuum chamber of the electrostatic mirror.}
	    \label{fig:manipulator}
\end{figure}

Based on the simulation results, we fabricated an aluminium plate with a 10- and 15-mm holes which was mounted on a magnetic transporter to serve as an adjustable collimator in the UHV chamber, as shown in Fig.~\ref{fig:manipulator}. In this way, the 10- and 15-mm collimators can be easily moved in front of the carbon foil of TD and switched as needed. A position sensitive micro-channel plate detector (MCP, Hamamatsu F1217-01) with delay-line readout \cite{Roentdek} was placed at the sample plate position to measure the beam spots and the muon rate. 
In many LEM experiments, magnetic fields $\le$~10 mT are applied parallel to the beam direction. Here,
we use a field of 10~mT to optimize the beam transport to the sample position. The \textbf{E×B} field of SR by rotating the muon spin from +10$^\circ$
to -10$^\circ$ direction, see Fig.~\ref{fig:LEM}.
Thus, while doing so-called transverse field measurements, i.e. magnetic field applied transverse to the muon spin direction, the background can be reduced to its lowest level. Results of these measurements are summarised in Table \ref{tab:15and12kVAtCenter} and show that, without a collimator, 42.6\% muons stop in the area of $10\times10$ mm$^2$, and only 16.5\% stop in the area of $5\times5$ mm$^2$ at V$_{mod}$ = 15 kV. While experiments on $10\times10$~mm$^2$ are still feasible with a 42\%-fraction of muons stopping in the sample, the fraction on the $5\times 5$~mm$^2$ sample makes LEM experiments impracticable.

\begin{figure}[!ht]
	    \centering
	    \includegraphics[width=1.0\textwidth]{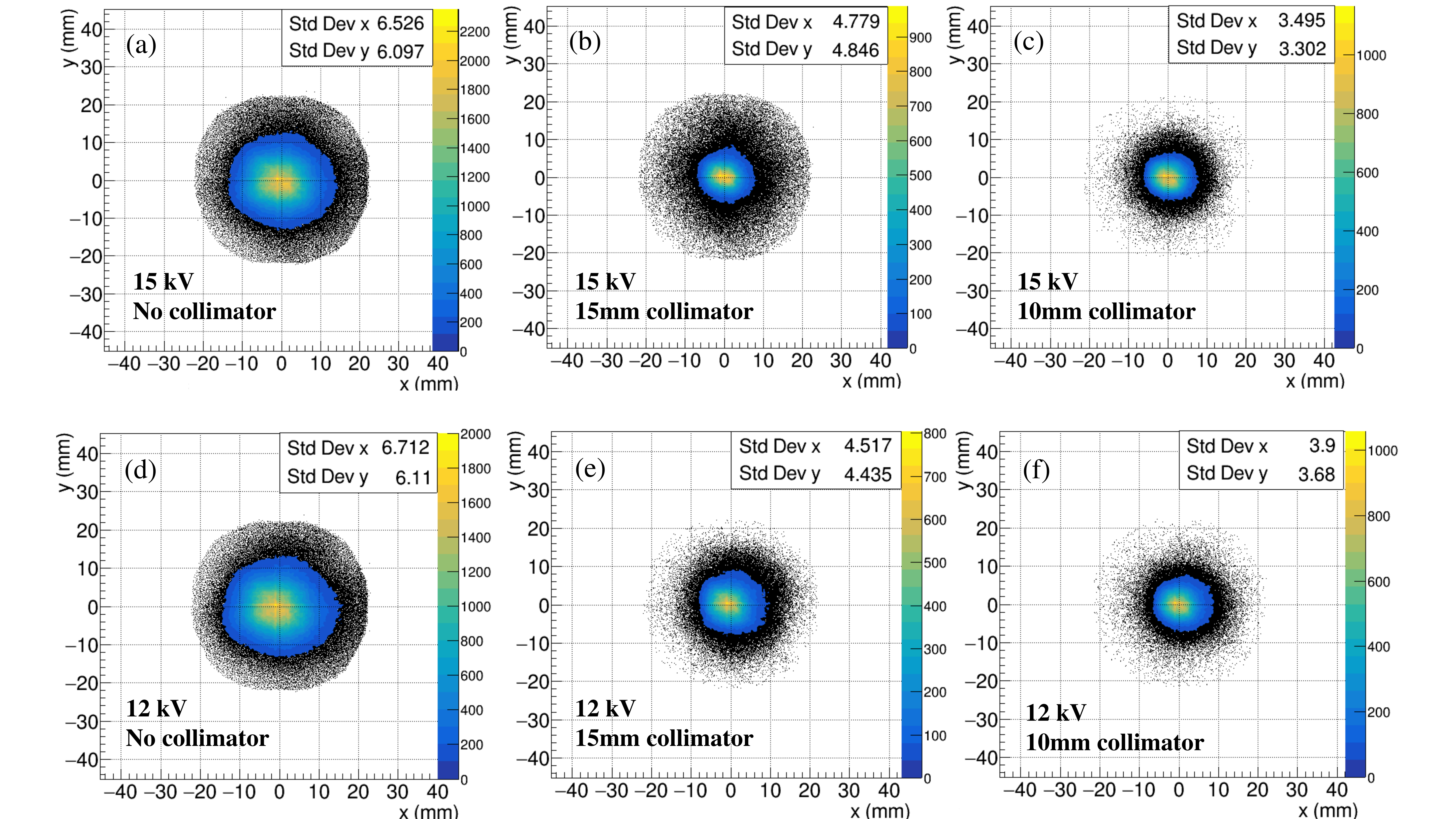}
	    \caption{Beam spots at the center with and without collimators at V$_{mod}$ = 15 and 12~kV. The black dots are background counts.}
	    \label{fig:BeamspotCenter}
\end{figure}

With a 15-mm collimator, the fractions of muons stopping in an area of $10\times10$ mm$^2$ and $5\times5$ mm$^2$ were significantly improved to 69.8\% and 39.2\%, respectively. The fractions increased even more with a 10-mm collimator at the expense of a further reduced LEM rate. 
The LEM rate is given by the coincidence of a detected decay positron with a 13-$\mu$s wide time window started by an event in the TD.
In general, with a 10- or 15-mm collimator, FoM (FoM = Fraction$^2$×LEM rate) was evidently improved compared to the ``No collimator'' setup. The 15-mm collimator is ideal for measurements on samples of $10\times10$ mm$^2$ and $5\times5$ mm$^2$ area at V$_{mod}$ = 15 kV. As shown in Table \ref{tab:15and12kVAtCenter}, the FoMs of small samples at V$_{mod}$ = 12 kV were also significantly enhanced with a 10- or 15-mm collimator. Figure~\ref{fig:BeamspotCenter} 
shows the measured beam spot at the MCP and its dependence of collimator size and transport energy. The RMS without collimator is approximately 6.6 mm. With a 15-mm collimator, the RMS is lowered to around 4.6 mm, while a 10-mm collimator further reduces the RMS to approximately 3.6 mm.


\begin{table}[!ht]
\centering
\caption{Comparison of beam spots with and without collimators at V$_{mod}$ = 15 and 12~kV. The LEM rate is normalized to 1~mA of proton current of the PSI proton accelerator HIPA \citep{Grillenberger_High_2021} used to generate the muons. FoM = Fraction$^2\times$~LEM rate.}
\begin{tabular}[htbp]{llllll}
\hline
Collimator diameter & {\begin{tabular}[l]{@{}l@{}}Fraction \\ 10×10\end{tabular}} & {\begin{tabular}[l]{@{}l@{}}Fraction \\ 5×5\end{tabular}}& {\begin{tabular}[l]{@{}l@{}}LEM Rate \\ (mAs)$^{-1}$ \end{tabular}} & {\begin{tabular}[l]{@{}l@{}}FoM  \\ 10×10\end{tabular}} & {\begin{tabular}[l]{@{}l@{}}FoM \\ 5×5\end{tabular}} \\
\hline
No collimator/15 kV & 0.426   & 0.165   & 1356.0   & 246.1  & 36.9   \\       
15 mm/15 kV         & 0.698   & 0.392   & 751.2    & 366.0  & 115.4  \\
10 mm/15 kV         & 0.816   & 0.455   & 477.3    & 317.8  & 98.8  \\
No collimator/12 kV & 0.388  & 0.148 & 1147.5  & 172.7  & 25.1   \\
15 mm/12 kV         & 0.655  & 0.315 & 580.0   & 248.8  & 57.6    \\
10 mm/12 kV         & 0.752  & 0.405 & 368.3   & 208.3  & 60.4    \\
\hline
\end{tabular}
\label{tab:15and12kVAtCenter}
\end{table}

\begin{figure}[!ht]
	    \centering
	    \includegraphics[width=1.0\textwidth]{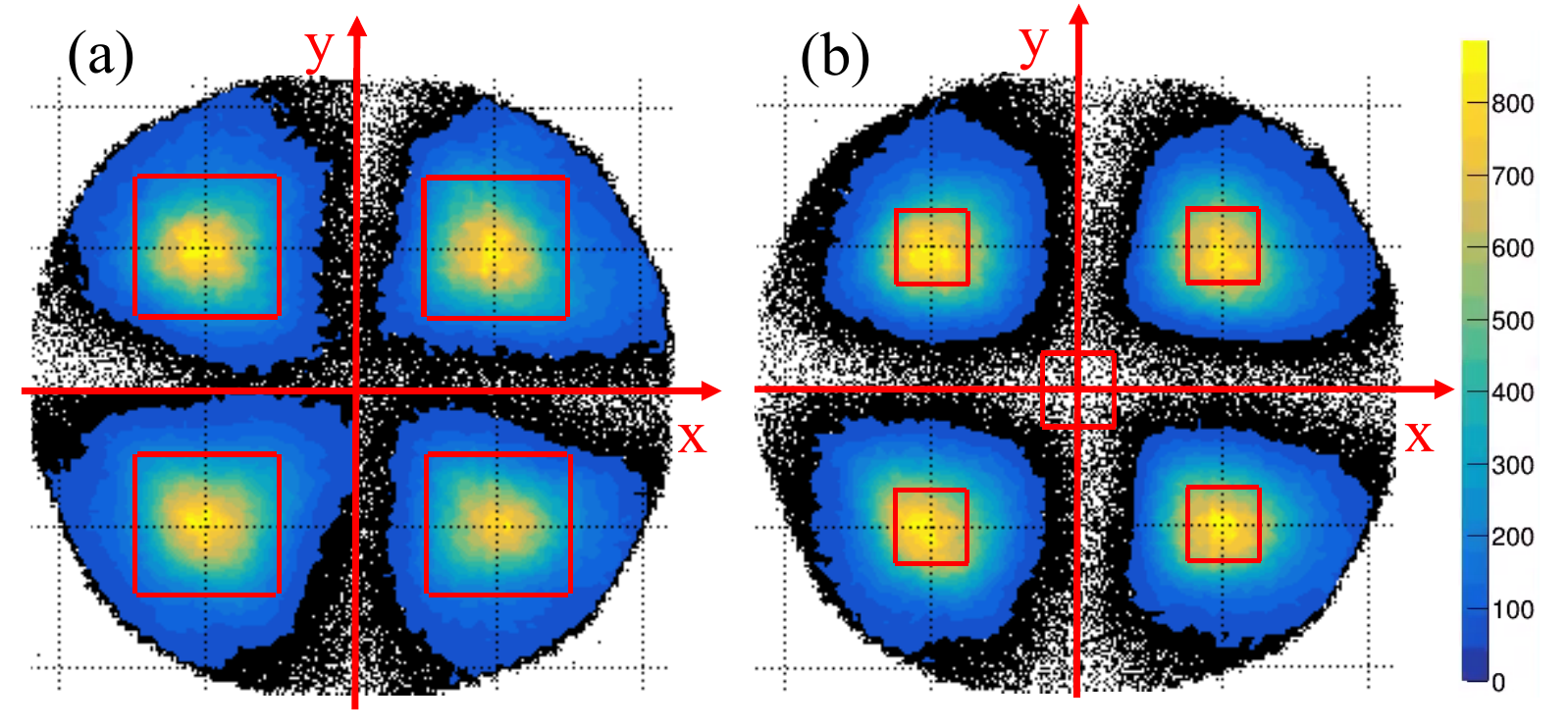}
	    \caption{Beam spots at four corners with (a) 15- and (b) 10-mm collimator at V$_{mod}$ = 15 kV. The red boxes in (a) and (b) show where the four $10\times10$ mm$^2$ and five $5\times5$ mm$^2$ samples will be placed, respectively. View in the beam direction. }
	    \label{fig:BeamspotFourCorners}
\end{figure}

Since the beam spots can be substantially reduced by the collimators, we considered mounting several small samples on one sample plate, e.g., four $10\times10$ mm$^2$ or five $5\times5$ mm$^2$ samples. By adjusting the voltages of the RA segments (RAL, RAR, RAT, RAB), the muon beam can be focused on different positions at the sample plate. As shown in Fig. \ref{fig:BeamspotFourCorners}, the beam can be centered at four corners of the MCP, left-top [-10 mm, 10 mm], right-top [10 mm, 10~mm], left-bottom [-10 mm, -10 mm], right-bottom [-10 mm, 10 mm]. With a 15-mm collimator, the beam spots are larger and more distorted than those with a 10-mm collimator, but they are still clearly separated from each other. Therefore, it is possible to put four $10\times10$ mm$^2$ samples at each of these four corners with a 15-mm collimator, and five $5\times5$ mm$^2$ samples (four corners and center) with a 10-mm collimator. Table \ref{tab:15and10mmFourcorners} summarizes the beam spot properties at four corners with 15- and 10-mm collimators with V$_{mod}$ = 15 kV.

\begin{table}[!ht]
\centering
\caption{Beam spot properties at four corners with a 15- and 10-mm collimator at V$_{mod}$ = 15 kV.}
\setlength{\tabcolsep}{2.0mm}
\begin{tabular}[htbp]{lllll}
\hline
Collimator size & Beam center position & {\begin{tabular}[l]{@{}l@{}}Fraction \\ 10×10\end{tabular}} & {\begin{tabular}[l]{@{}l@{}}LEM Rate \\ (mAs)$^{-1}$\end{tabular}} & {\begin{tabular}[l]{@{}l@{}}FoM  \\ 10×10\end{tabular}} \\
\hline
15 mm & Left top [-10, 10]     & 0.689 & 725.2 & 344.3  \\
15 mm & Right top [10, 10]     & 0.684 & 740.4 & 346.4  \\
15 mm & Left bottom [-10, -10] & 0.659 & 739.2 & 321.0  \\
15 mm & Right bottom [-10, 10] & 0.690 & 743.8 & 354.1  \\
10 mm & Left top [-10, 10]     & 0.365 & 470.9 & 62.7  \\
10 mm & Right top [10, 10]     & 0.351 & 474.8 & 58.5  \\
10 mm & Left bottom [-10, -10] & 0.352 & 470.0 & 58.2  \\
10 mm & Right bottom [-10, 10] & 0.371 & 473.3 & 65.1  \\
\hline
\end{tabular}
\label{tab:15and10mmFourcorners}
\end{table}


In the next section we show how the collimators and beam steering can be used to carry out $\mu$SR experiments on four or five small samples mounted simultaneously on one sample plate. 

\section{$\mu$SR measurements}
\label{sec:muSR}
\subsection{One $10\times10$ mm$^2$ sample at the center}
As an example of usefulness of a collimator for small samples mounted at the center of a sample plate, we compare the FoM of a $10\times10$ mm$^2$ area measured with and without the 15-mm collimator. We mounted a 20~nm SrTiO$_3$/10 nm Nd$_{0.8}$Sr$_{0.2}$NiO$_2$/SrTiO$_3$ sample with an area of $10\times10$ mm$^2$ on the sample plate and measured it using beam-spot-optimized 12 kV transport settings with 4.5 keV implantion energy. This material is interesting due to the recent observation of superconductivity in Nd$_{0.8}$Sr$_{0.2}$NiO$_2$ which ushered in a new class of layered nickelate superconductors \citep{li2019superconductivity}. These samples can be grown only in thin films and on small substrates of $10\times10$~mm$^2$ or $5\times5$~mm$^2$ size. Since the top layers of the sample are thin, one needs a low muon implantation energy to stop a maximum fraction of the muons in the nickelate layer \citep{fowlie2022intrinsic}. As shown in Table \ref{tab:STO_NSNO_STO}, the FoM with 15-mm collimator is about 20\% larger than that without collimator. The use of the collimator with a corresponding increase of asymmetry and reduction of background turned out to be crucial for the analysis of the $\mu$SR data in zero magnetic field which were used to determine the details of the intrinsic magnetic state as a function of sample doping \citep{fowlie2022intrinsic}.
Note that the LEM rates are lower than those measured with the MCP in Tab.~\ref{tab:15and12kVAtCenter} because, in comparison to the MCP setup, more positrons from muon decay are stopping in the radiation shield and other components of the sample cryostat.

\begin{table}[!ht]
\centering
\caption{FoMs of $10\times10$ mm$^2$ SrTiO$_3$/Nd$_{0.8}$Sr$_{0.2}$NiO$_2$/SrTiO$_3$ sample with and without collimator.}
\setlength{\tabcolsep}{5mm}
\begin{tabular}[htbp]{llll}
\hline
Collimator diameter &  Asymmetry A & {\begin{tabular}[l]{@{}l@{}}LEM rate R \\ (mAs)$^{-1}$\end{tabular}} & {\begin{tabular}[l]{@{}l@{}}FoM  \\ A$^2$×R\end{tabular}} \\
\hline
No Collimator  &  0.056 &  905  &  2.8  \\
15 mm          &  0.092 &  390  &  3.3  \\
\hline
\end{tabular}
\label{tab:STO_NSNO_STO}
\end{table}

\subsection{Four $10\times10$ mm$^2$ SrTiO$_3$ samples at four corners}
\label{subsec:FourSTO}
\begin{figure}[!ht]
	    \centering
	    \includegraphics[width=1.0\textwidth]{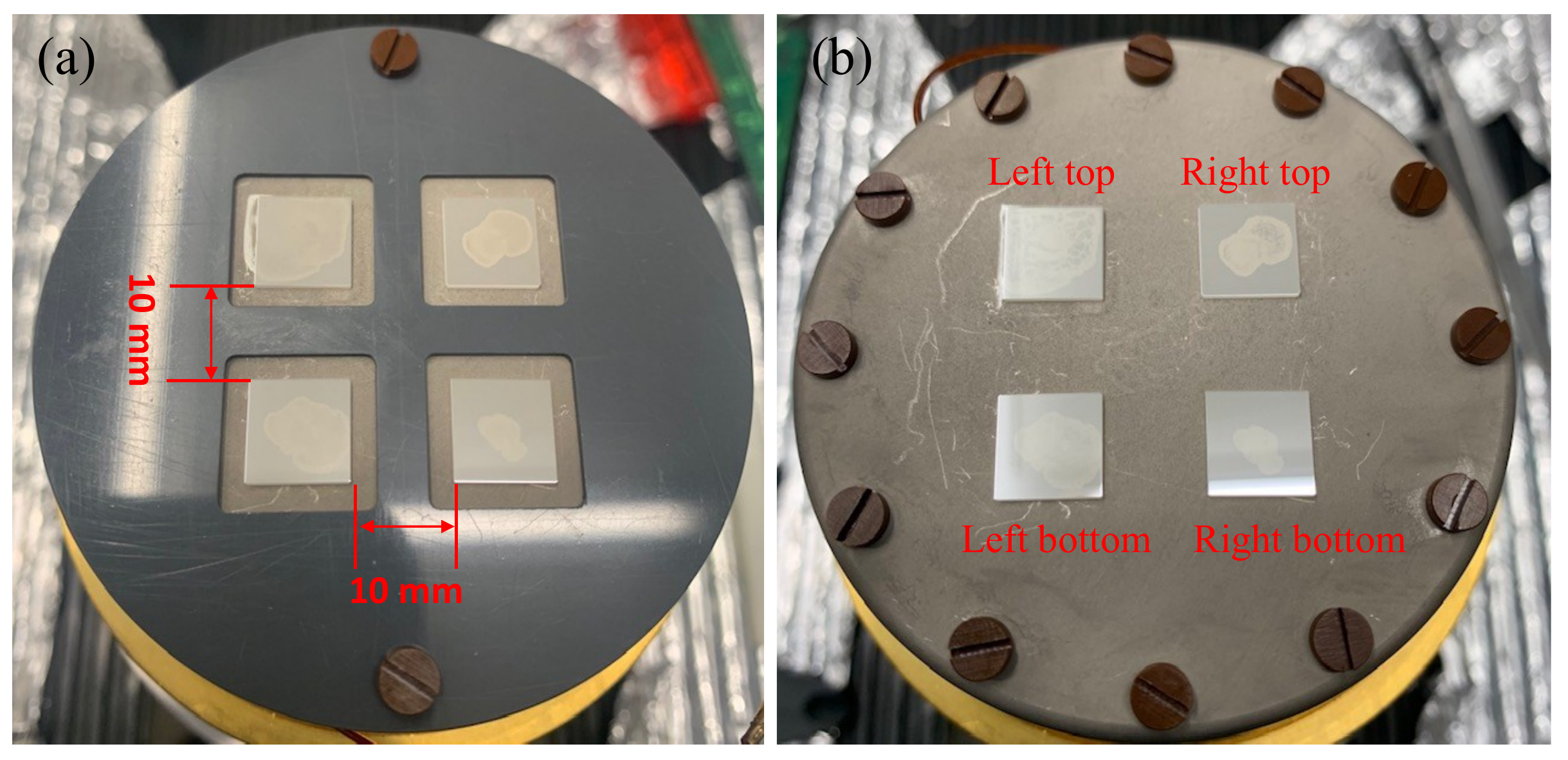}
	    \caption{Four $10\times10$ mm$^2$ SrTiO$_3$ samples at four corners of one sample plate. The mask (a) was designed to help mounting samples at the exact positions. View in the muon beam direction.}
	    \label{fig:FourSTO}
\end{figure}

\begin{figure}[!ht]
	    \centering
	    \includegraphics[width=1.0\textwidth]{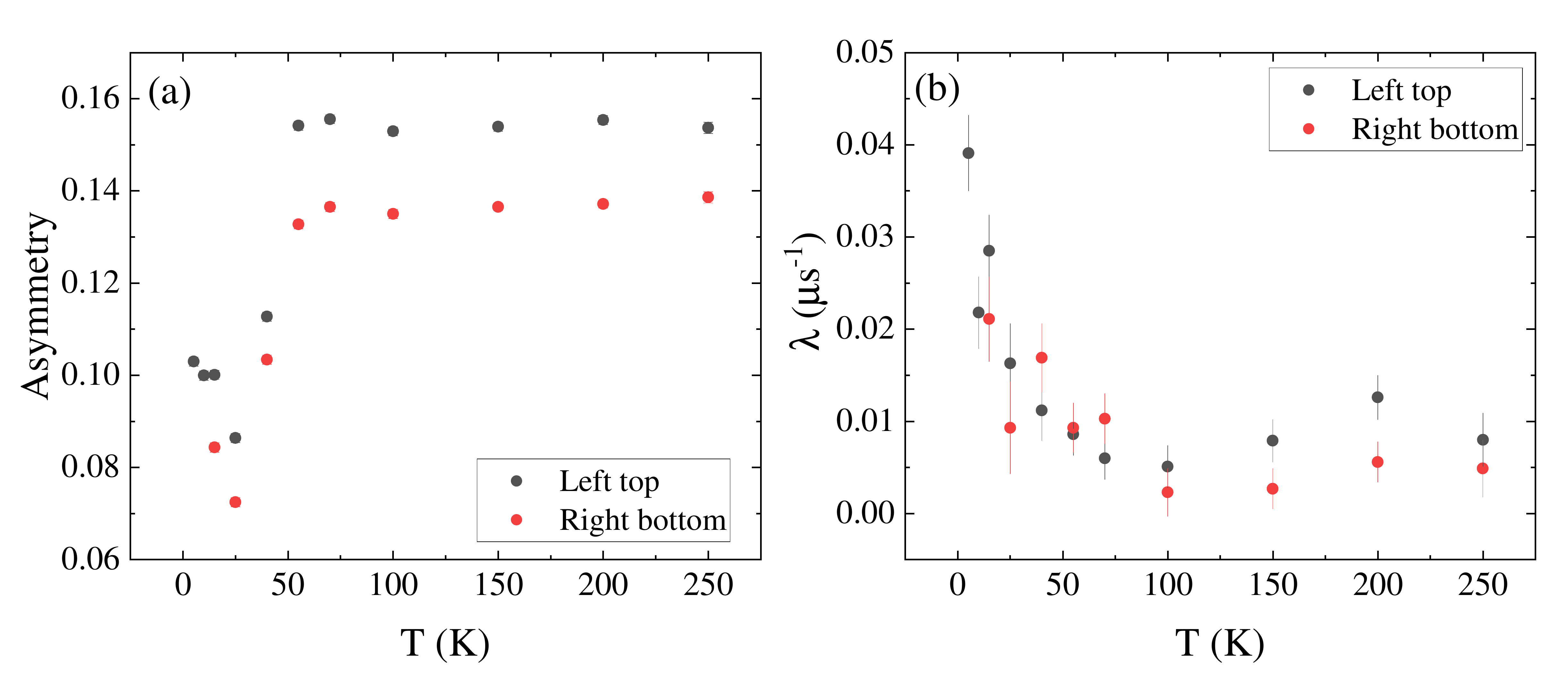}
	    \caption{Asymmetries and relaxation rates of $10\times10$ mm$^2$ SrTiO$_3$ samples at left-top [-10 mm, 10 mm] (black circles) and right-bottom [-10 mm, 10 mm] (red circles) of sample plate as a function of temperature.}
	    \label{fig:MuSR_FourSTO}
\end{figure}

As shown in Fig. \ref{fig:FourSTO}, we designed a mask with four quadratic openings that covers the sample plate and is used to accurately and reproducibly mount the samples in well defined positions. Here we used SrTiO$_3$ samples which is an excellent reference sample that has a full diamagnetic muon asymmetry at room temperature (0.23 in LEM). We measured the asymmetries and relaxation rates of the samples at the left-top [-10 mm, 10 mm] and right-bottom [-10 mm, 10 mm] as a function of temperature. These measurements were performed using 15 kV transport settings optimized by MCP with a 15-mm collimator, 10 mT magnetic field anti-parallel to the beam direction in the sample region, and 16.2 keV muon implantion energy (i.e. with -4 kV applied to the sample plate). As shown in Fig. \ref{fig:MuSR_FourSTO}, the asymmetry measured in both sample exhibit similar behavior, and their relaxation rates are nearly identical. From 250 to 70 K, the asymmetry remains constant. However below 70 K, it quickly decreases due to the formation of a shallow muonium state, which reduces the observable asymmetry \citep{salman2014direct}. When the temperature is below 25 K, the asymmetry increases again. This may be caused by the reduction of the muonium formation probability due to the increasing dielectric constant of SrTiO$_3$ \citep{salman2014direct}.

From the MCP measurements in Table \ref{tab:15and10mmFourcorners}, the stopped muons' fractions in the $10\times10$~mm$^2$ area at these positions are about 69\%. Therefore, the asymmetries should be 0.159 (0.23$\times$0.69). From Fig. \ref{fig:MuSR_FourSTO} and Table \ref{tab:15mmFourcorners_Asy}, the sample at the left top matches very well with what we expect from the MCP measurement. However, those at the other three positions have about 4\% - 10\% lower asymmetries. This could be because the active area of MCP has a diameter of 42 mm, while the sample plate has a diameter of 70 mm. Consequently, the beam spots at these four corners may extend outside the MCP's active region, causing the fractions in the $10\times10$ mm$^2$ region to be slightly different between the MCP and $\mu$SR measurements. 

\begin{table}[!ht]
\centering
\caption{Asymmetries of four $10\times10$ mm$^2$ SrTiO$_3$ samples mounted on one plate and measured at 250~K.}
\setlength{\tabcolsep}{3.5mm}
\begin{tabular}[htbp]{lllll}
\hline
& \begin{tabular}[c]{@{}l@{}}Left top\\ {[}-10,10{]}\end{tabular} & \begin{tabular}[c]{@{}l@{}}Right top\\ {[}10,10{]}\end{tabular} & \begin{tabular}[c]{@{}l@{}}Left bottom\\ {[}-10,-10{]}\end{tabular} & \begin{tabular}[c]{@{}l@{}}Right bottom\\ {[}-10,10{]}\end{tabular} \\
\hline
Asymmetry     & 0.154 & 0.148 & 0.138 & 0.144 \\
\hline
\end{tabular}
\label{tab:15mmFourcorners_Asy}
\end{table}

\subsection{Five $5\times5$ mm$^2$ SrTiO$_3$ samples on one sample plate}
\begin{figure}[!ht]
	    \centering
	    \includegraphics[width=1.0\textwidth]{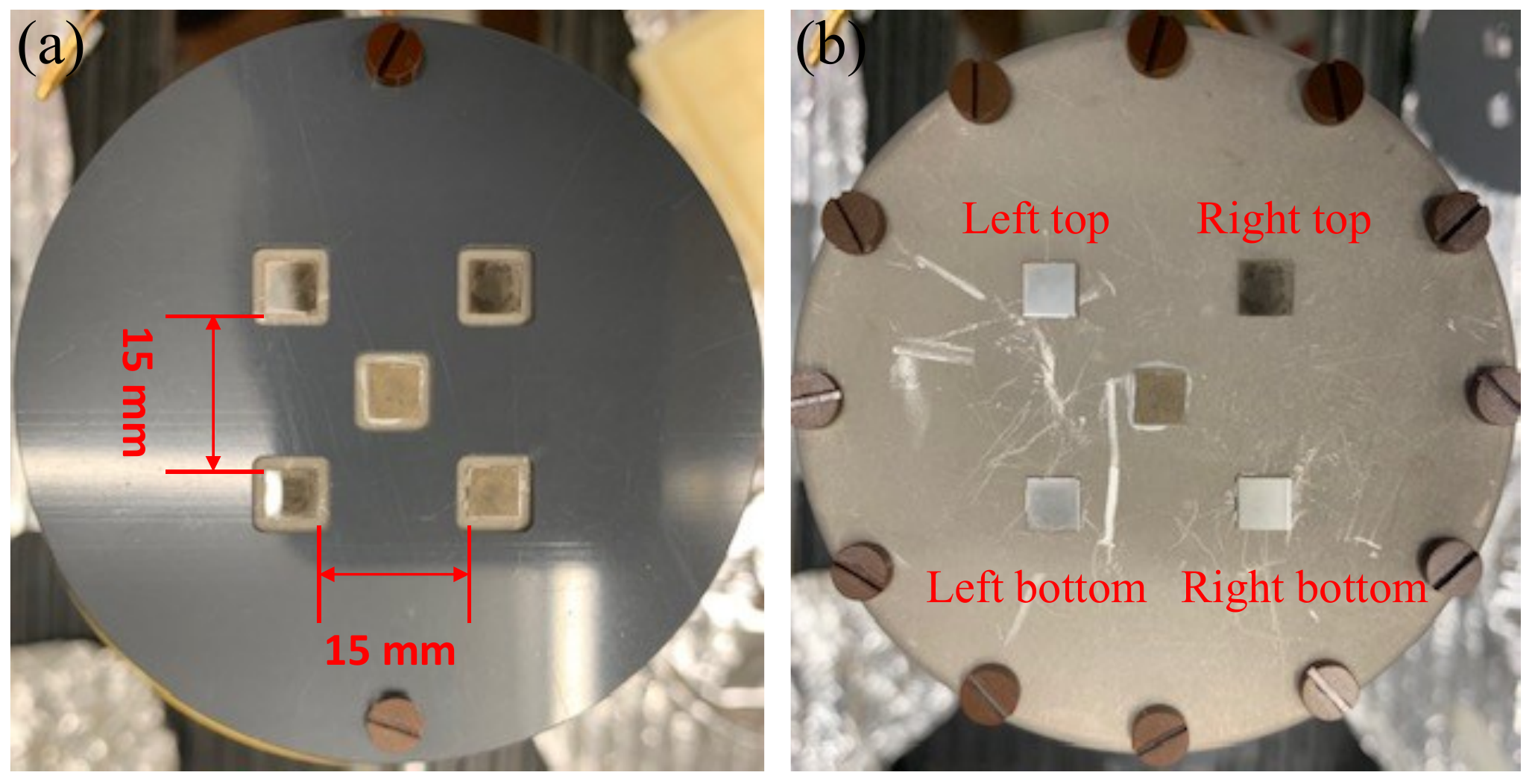}
	    \caption{Five $5\times5$ mm$^2$ SrTiO$_3$ samples on one sample plate. One is at the center and the others are at the four corners.}
	    \label{fig:FiveSTO}
\end{figure}

\begin{figure}[!ht]
	    \centering
	    \includegraphics[width=1.0\textwidth]{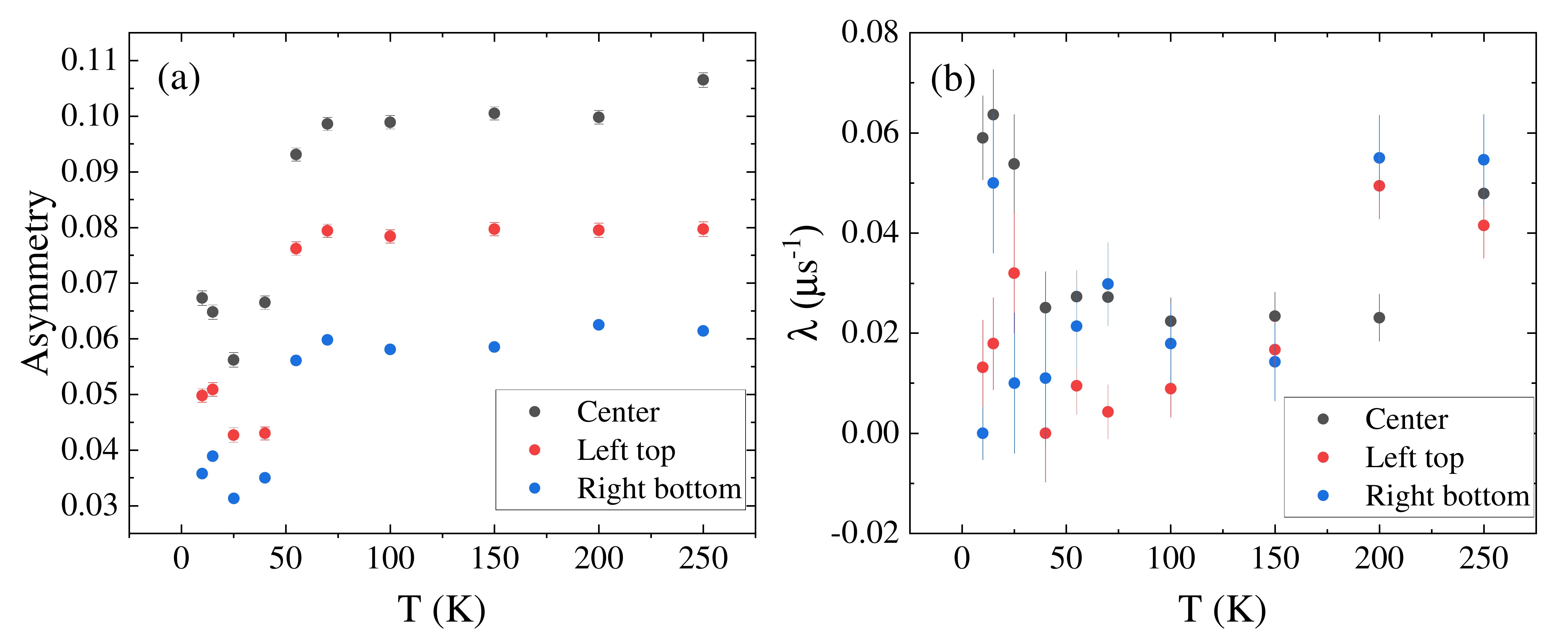}
	    \caption{Asymmetries and relaxation rates of $5\times5$ mm$^2$ SrTiO$_3$ samples at the center (black circles), left-top [-10 mm, 10 mm] (red circles) and right-bottom [-10 mm, 10 mm] (blue circles) of sample plate as a function of temperature.}
	    \label{fig:MuSR_FiveSTO}
\end{figure}

Measurements on five $5\times5$ mm$^2$ SrTiO$_3$ samples were performed, in a similar fashion to the measurements in Section \ref{subsec:FourSTO}, except that the 10~mm collimator was utilized here. As shown in Fig. \ref{fig:FiveSTO}, we also designed a mask to mount five $5\times5$ mm$^2$ samples. The $\mu$SR results in Fig. \ref{fig:MuSR_FiveSTO} and Table \ref{tab:10mmFivecorners_Asy} show, that the asymmetry of the sample at the center is 20\% - 40\% higher than those at the other four corner positions, and that the asymmetry in the center is in good agreement with MCP measurements, where we expect an asymmetry of 0.105 (0.23$\times$0.455). The beam spots at the corners are more distorted than that at the center, causing the reduction of asymmetry due to the smaller fraction of muons stopping on the sample. Nevertheless, the small sample measurements clearly show the characteristic temperature dependencies of the asymmetry and relaxation rate of SrTiO$_3$ due to shallow muonium formation below 70~K.

\begin{table}[!ht]
\centering
\caption{Asymmetries of five $5\times5$ mm$^2$ SrTiO$_3$ samples mounted on one plate and measured at 250~K.}
\setlength{\tabcolsep}{1.8mm}
\begin{tabular}[htbp]{llllll}
\hline
& Center & \begin{tabular}[c]{@{}l@{}}Left top\\ {[}-10,10{]}\end{tabular} & \begin{tabular}[c]{@{}l@{}}Right top\\ {[}10,10{]}\end{tabular} & \begin{tabular}[c]{@{}l@{}}Left bottom\\ {[}-10,-10{]}\end{tabular} & \begin{tabular}[c]{@{}l@{}}Right bottom\\ {[}-10,10{]}\end{tabular} \\
\hline
Asymmetry     & 0.107 & 0.080 & 0.061 & 0.069 & 0.061 \\
\hline
\end{tabular}
\label{tab:10mmFivecorners_Asy}
\end{table}

Although the use of the collimators leads to a significant reduction of the muon beam rate, i.e. the total number $N$ of accumulated muon decay events in a given time $\Delta T_{acc}$, measurements on small samples can be carried out using the same data collection time $\Delta T_{acc}$ with an even better statistical precision, which is determined by $FoM = A^2\times N$. 
This is because the gain in asymmetry squared ($A^2$) overcompensates the loss in total events $N$.

\section{Summary}
\label{sec:Summary}
By beam collimation and tuning, we have demonstrated the feasibility to carry out LE-$\mu$SR measurements on samples as small as $5\times5$ mm$^2$ using the LEM facility at PSI. This is an important improvement towards the utilization of low-energy muon beams, which suffer from a large phase space and relatively low rates, limiting measurements to larger samples (typically $25\times25$ mm$^2$)
This development is significant not only for many magnetic and superconducting materials -- which often can be grown only on small substrates -- but also for research using external stimuli such as application of electric fields and illumination. For electric field measurements, small samples offer the benefit of a better bias control over the smaller sample area. For illumination experiments, a small sample allows better light focusing, which improves the illumination effect by reaching higher uniformity and intensity.

Using collimators to reduce the beam spot size, four $10\times10$ mm$^2$ or five $5\times5$ mm$^2$ samples can be mounted simultaneously on a sample plate. Thus, avoiding loss of beam time for sample changes -- which usually takes three hours, hence enhancing the utilization efficiency of the LEM facility with a typical oversubscription of two. 

Using the collimator reduces the LEM rate, but this can be compensated. The installation of a slanted muon target E in 2019 has already resulted in a 40\% increase in the surface $\mu^+$ beam intensity in the $\mu$E4 beamline \citep{berg_target_2016,Daniela2022}. Our recent work shows that by replacing the last quadrupole triplet of the $\mu$E4 beamline by a special solenoid magnet with minimized fringe fields, one can obtain an additional increase of 40\% of the original beam intensity at the LEM moderator \citep{zhou2022simulation}. The installation of this solenoid is foreseen for 2025. Both improvements will result in a factor of two increase in LEM intensity. In a longer term perspective, the High-Intensity Muon Beams (HIMB) project will provide a surface muon beam rate of about twenty times the current $\mu$E4 rate \citep{aiba2021science}. If approved, this project will be realized in 2028.

\section{Declaration of competing interest}
The authors declare that they have no known competing financial interests or personal relationships that could have appeared to influence the work reported in this paper.

\section{Data availability}
Data will be available on request.

\section{Acknowledgments}
All the measurements have been performed at the Swiss Muon Source S$\mu$S, Paul Scherrer Institute, Villigen, Switzerland. We are grateful to Hans-Peter Weber for his excellent technical support.

\bibliographystyle{elsarticle-num-names} 
\bibliography{References}

\end{document}